# Spanish philosophers' perceptions of pay to publish and open access: books versus journals, more than a financial dilemma

Ramón A. Feenstra[1]  & Emilio Delgado López-Cózar[2]

[1]Universitat Jaume I de Castelló (Spain)  [2]Universidad de Granada (Spain)

**Abstract**

This study examines habits and perceptions related to pay to publish and open access practices in fields that have attracted little research to date: philosophy and ethics. The study is undertaken in the Spanish context, where the culture of publication and the book and journal publishing industry has some specific characteristics with regard to paying to publish, such as not offering open access distribution of books published for a fee. The study draws on data from a survey of 201 researchers, a public debate with 26 researchers, and 14 in-depth interviews. The results reveal some interesting insights on the criteria researchers apply when selecting publishers and journals for their work, the extent of paying to publish – widespread in the case of books and modest for journals– and the debates that arise over the effects it has on manuscript review and unequal access to resources to cover publication fees. Data on the extent of open access and the researchers' views on dissemination of publicly funded research are also presented.

Ramón A. Feenstra     Emilio Delgado López-Cózar
feenstra@uji.es       edelgado@ugr.es

> **Key points:**
>
> - Paying to publish is favoured in the case of book publications. The publication of books in the local language is defended as culturally and socially valuable. Payment to publish books does not imply open access distribution.
>
> - Paying to publish raises questions about manuscript review processes, their quality and equal opportunities.
>
> - Open access is commonplace in Spanish philosophy because many Spanish academic journals are financed with public funds.
>
> - Most of the researchers surveyed agree that publicly funded research should be openly disseminated without restrictions. Note that open access is understood as a model that is free for the reader and the author, and is financed with public funds.

Since its introduction, open access (OA) of scientific publications has generated debate on how it should be paid for (Harnad et al., 2004; Truth, 2012; Björk & Solomon, 2015; Bo-Christer Björk, 2017; Severin et al., 2018; Edelmann & Schoßböck, 2020). The traditional publication model, financed through subscriptions, was free for the author, but paid for (and with limited access) by readers. However, an alternative form of publishing has recently become established within the OA model, based on article processing charges (APC) paid by the author (or their institution) and available to the reader without charge (Schroter & Tite, 2006; Gadagkar, 2016; Leopold, 2014; Araiza-Díaz, Ramírez-Godoy & Díaz-Escoto, 2019). In sum, the business model has shifted from pay-to-read to pay-to-publish.

An intense debate has arisen over the implications of the pay-to-publish model, which has mainly been implemented and studied in regard to research journals (Gadagkar, 2016). Some authors consider that this business model can affect the quality of published material, can give rise to possible conflicts of interest in journals, or can distort review processes (Mabe, 2004; van Dalen, 2013; Leopold, 2014; Gadagkar, 2016; Al-Khatib & da Silva, 2017). Criticism has also been levelled at the disproportionate profits accrued from the excessive fees (compared to the cost of production) charged to authors (Buranyi, 2017; Pinter, 2018; Dal-Re, 2019), which create inequalities among researchers, who do not all have the same access to funding for these payments (Mabe, 2004; Gadagkar, 2008; van Dalen 2013; Leopold 2014; Tzarnas & Tzarnas, 2015; Al-Khatib & da Silva, 2017). This is especially the case for young researchers, who often identify the high cost of APCs as one of the main problems of OA publishing (Nicholas et al., 2019; Jamali et al., 2020; Rodríguez-Bravo & Nicholas, 2020; Nicholas et al., 2020). Other authors warn that behind the attractive OA label lie some 'predatory' journals, known for their tendency to extort high fees from authors (Beall, 2012; Truth, 2012; Shehata & Elgllab, 2018; Tzarnas & Tzarnas, 2015; Al-Khatib & da Silva, 2017; Abad-García, 2019; Nicholas et al., 2019a; Alonso-Arévalo, Saraiva & Flórez-Holguín, 2020).

The option to pay to publish for OA is also available in book publishing through book processing charges (BPC) (Eve, 2014). This model is not as widespread as among journals, although it is growing (Ferwerda, Pinter & Stern, 2017; Giménez Toledo & Córdoba Restrepo, 2018; Capaccioni, 2020). The BPC model has also sparked debate about the effects on manuscript review (Knöchelmann, 2018) and has met with a low level of acceptance in the humanities and social sciences. This rejection is due above all to the high cost and scarce funding opportunities (Giménez-Toledo, 2018; Jobmann & Schönfelder, 2019). Humanities publishers are also concerned that this model may not be sustainable if researchers cannot meet the costs involved (Severin et al., 2018). Other analysts have also documented huge disparities among BPC fees (Jubb, 2017; Ferwerda, Pinter & Stern, 2017). Finally, other initiatives and studies have explored alternative funding formats to stimulate open books, both in general (Reinsfelder & Pike, 2018; Barnes & Gatti, 2019) and in the humanities (Snijder, 2019; Jobmann & Schönfelder, 2019; Eve et al., 2017).

The present paper examines the habits and perceptions related to pay to publish and OA in the fields of Spanish philosophy and ethics. This research is particularly important because it focuses on disciplines that have received scant academic attention, and in which the publication of books in the vernacular language addressed to local readerships is an essential publication format (Fry, 2009; Hammarfelf & Rijcke, 2015; Hammarfelf, 2017; Engels et al., 2018; Kulczycki, & Korytkowski, 2019). Moreover, the culture of academic publishing in Spanish philosophy is idiosyncratic in that payments to publish books generally go towards subsidising a precarious publishing market, and unlike the rationale behind the BPCs, payment is requested to ensure publication is viable or possible. In this case, payment is not used to provide OA distribution of the book; rather, the author buys a stipulated number of books to guarantee the publisher's production costs are covered. This somewhat unconventional practice has attracted little research attention (Giménez-Toledo, Tejada-Artigas & Borges-De-Oliveira, 2019). In turn, the Spanish journal publishing sector in the field of philosophy is supported by a model of public financing in which authors do not pay to publish and readers are not charged to access journal content. In other words, OA is subsidised through public funds that support academic journals affiliated to universities and research centres, both in the humanities in general (Melero, 2017; Claudio-González & Villarroya, 2017) and philosophy in particular (Feenstra & Pallarés-Domínguez, 2021). This model has not, however, taken hold in this context for book publication (López-Carreño, Delgado-Vázquez & Martínez-Méndez, 2021).

Several studies have analysed and gathered data on OA in the Spanish context, whether in reference to a particular institution (Serrano-Vicente, Melero & Abadal, 2016), the whole country (Bernal, 2010; Hernández-Borges et al., 2006; Ruiz-Pérez & Delgado-López-Cózar, 2017; Segado-Boj et al, 2018), or specific groups such as early career researchers (Rodríguez-Bravo & Nicholas, 2019; Rodríguez-Bravo & Nicholas, 2020). However, few studies examine the perceptions and practices adopted in the areas of the humanities in Spain (Serrano-Vicente, Melero & Abadal, 2016; Ruiz-Pérez & Delgado-López-Cózar, 2017) and there is no specific research in the field of philosophy, the focus of the present study.

In sum, to our knowledge no previous studies have specifically examined the views of philosophy researchers on the questions of paying to publish and OA. This study aims to fill this gap by finding out the extent to which these researchers accept, engage with or challenge these practices, and document their dilemmas, reticences or intellectual misgivings.

**Methodology**

In this study we apply triangulation methodology comprising a self-administered questionnaire, a debate held at the annual meeting of scientific association, and 14 in-depth interviews. These three data-gathering techniques yielded quantitative and qualitative information about pay to publish and OA.

*Self-administered questionnaire*

The study population comprised university researchers and faculty working in the knowledge areas of philosophy and ethics in Spain, together with researchers from the Institute of Philosophy at the Spanish National Research Council (CSIC). We identified the members of this academic community through a systematic search of the websites of Spanish universities. Through these inquiries we identified 541 faculty members and researchers, of whom 521 worked in universities and 20 in the CSIC; 44 universities (37 public and seven private) took part in the study and responses were received from all but three institutions. Table 1 shows the distribution by knowledge area.

**Table 1**
**Demographics of the survey of Spanish university faculty and researchers in philosophy and ethics**

| ÁREA DE CONOCIMIENTO | Población | Muestra | Tasa de respuesta |
|---|---|---|---|
| Philosophy | 380 | 115 | 30.5% |
| Ethics | 161 | 86 | 52.8% |
| Total | 541 | 201 | 37.0% |

The online survey remained open for responses between February and June 2019. On 25 February, a message was sent to the institutional email address of the 541 faculty members and researchers identified, followed by two reminders. The survey was also promoted by three Spanish philosophy associations: the Spanish Association for Ethics and Political Philosophy (AEEFP), the Academic Society of Philosophy (SAF) and the Spanish Philosophy Network (REF). The survey was closed on 14 June 2019.

The data was collected using Google Forms. The questions were divided into four main sections: 1) information search behaviour, 2) scientific evaluation, 3) ethics in scientific publication and 4) communication practices, the latter being the main focus of the present study.

The questions for the communication practices section were as follows:

- Q1. Please evaluate the importance of the following factors when selecting a journal or publisher for your work on a scale of 1 to 5, where 1 = Not important at all, and 5 = Very important.
    - Open access publication
    - Impact measured by citation countsPrestige: the tradition of the journal or publisher
    - Speed of publication
    - Quality of manuscript selection and peer review process
    - Ease of access to the publication's editors
    - Publisher's subject orientation or specialisation
    - Journal's dissemination and visibility in databases
    - No publication fee charged

- Q2. Has a publisher that has published your work ever asked you to pay to publish a book (whether an individual or a collective publication)? Please select one of the following options:
    - Never
    - Only rarely
    - Not often
    - Often
    - Always
- Q3. If the answer to the previous question is yes, how much did the publishers ask you to pay? Please select one of the following options:
    - Less than 500 euros
    - 500 to 1,000 euros
    - 1,001 to 2,000 euros
    - 2,001 to 3,000 euros
    - More than 3,000 euros
- Q4. Has a journal that has published your work ever asked you to pay to publish an article? Please select one of the following options:
    - Never
    - Only rarely
    - Not often
    - Often
    - Always
- Q5. If the answer to the previous question is yes, how much did the journal ask you to pay? Please select one of the following options:
    - Less than 500 euros
    - 500 to 1,000 euros
    - 1,001 to 2,000 euros
    - 2,001 to 3,000 euros
    - More than 3,000 euros
- Q6. Of your last ten publications, how many were published in open access format?
- Q7. Do you agree with the following statement? "All publications by researchers who receive public funds should be published in open access format, without exception".

The questionnaire included an open question where respondents could freely express their opinions on the study topic.

*Debate at the annual meeting of a scientific society*

On 30 May 2019, the authors of this study participated in a debate during the annual meeting of the Spanish Association for Ethics and Political Philosophy (AEEFP), held in the Institute of Philosophy at the CSIC (Madrid). The debate lasted 65 minutes and contributions to the discussion from 26 researchers. This qualitative information came from researchers working in a number of different universities and with diverse academic careers: six research fellows (23.1%), three lecturers (11.5%), ten senior lecturers (38.5%), and seven full professors (26.9%). Their contributions provided valuable input in preparing the in-depth interviews, the final stage of the research.

*Interviews*

The interviews took place in September and October 2019. The 14 interviewees were selected according to the criteria of affiliation, professional category, gender and disciplinary area in order to guarantee the widest possible range of profiles. Seven of the interviewees were men and seven were women, and seven worked in the field of ethics and seven in philosophy. The interviewees were affiliated to the universities of Barcelona, Castellón, Complutense (Madrid), Granada, Murcia, Valencia, Zaragoza and the Basque Country, as well as the Institute of Philosophy at the CSIC. The semi-structured interviews lasted an average of 35.30 minutes; the shortest was 14.14 minutes and the longest, 59.10 minutes. The interviews were then transcribed for analysis. Interviewees were asked about their experiences and perceptions of paying to publish and OA.

**Table 2. List of university faculty and researchers in philosophy and ethics interviewed according to career position.**

| Nº | Career position | Area | Code |
|---|---|---|---|
| 1 | Research fellows | E | RfE1-i |
| 2 | Research fellows | E | RfE2-i |
| 3 | Research fellows | PH | RfPh1-i |
| 4 | Lecturer | PH | LPh1-i |
| 5 | Lecturer | PH | LPh2-i |
| 6 | Senior lecturer | PH | SlPh1-i |
| 7 | Senior lecturer | PH | SlPh2-i |
| 8 | Senior lecturer | PH | SlPh3-i |
| 9 | Senior lecturer | E | SlE1-i |
| 10 | Senior lecturer | E | SlE2-i |
| 11 | Professor | PH | PPh1-i |
| 12 | Professor | E | PE1-i |
| 13 | Professor | E | PE2-i |
| 14 | Professor | E | PE3-i |

**Table 3. List of acronyms**

| Stage career | |
|---|---|
| Research Fellow | Rf |
| Post-doctoral researcher | Pdr |
| Lecturer | L |
| Senior Lecturer | Sl |
| Professor | P |
| Area | |
| Ethics | E |
| Philosophy | Ph |
| Source of information | |
| Interview | -i |
| Open section of the survey | -s |

**Results**

**The importance of paying to publish and open access in researchers' journal and publisher choices**

Before examining Spanish philosophers' pay to publish and OA practices, we wanted to find out how important they consider these factors when selecting a journal or publisher for their work. The results showed that Spanish philosophy and ethics researchers greatly value the absence of publication fees, which they rate as the third most relevant criterion in their decision (Figure 1). Only the prestige of the journal or publisher and its subject specialisation are considered more important when choosing where to submit their work for publication. In contrast, OA is not a highly relevant factor, lying second from the bottom of the table.

**Figure 1. Criteria used in selecting journals and publishers in which to publish, according to university lecturers and researchers in philosophy and ethics in Spanish institutions**

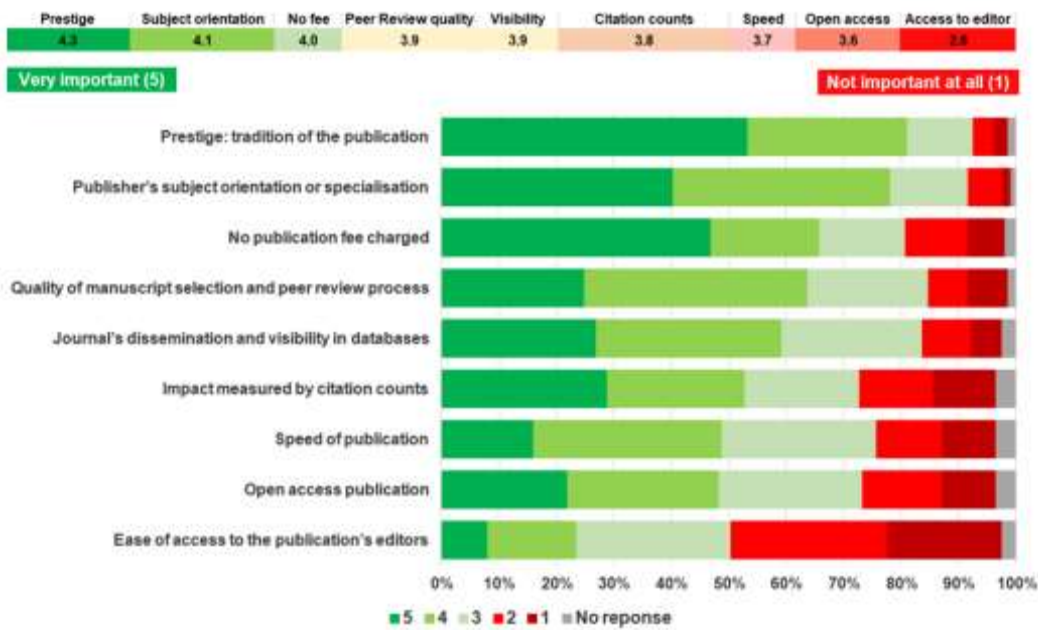

**Perceptions of pay to publish and associated practices in Spanish philosophy: books versus journals**

Philosophy and ethics researchers adopt different pay-to-publish practices, depending on whether they want to publish an article or a book. In the former, payment is not common practice as 80% of the survey respondents reported never having paid to publish a journal article (Figure 2), compared to only 37% in the case of publishing a book. Notably, only 2% of respondents said they often or always paid to have an article published, compared to 30% in the case of books.

Figure 2. Extent of pay-to-publish practices among Spanish university philosophy and ethics lecturers and researchers

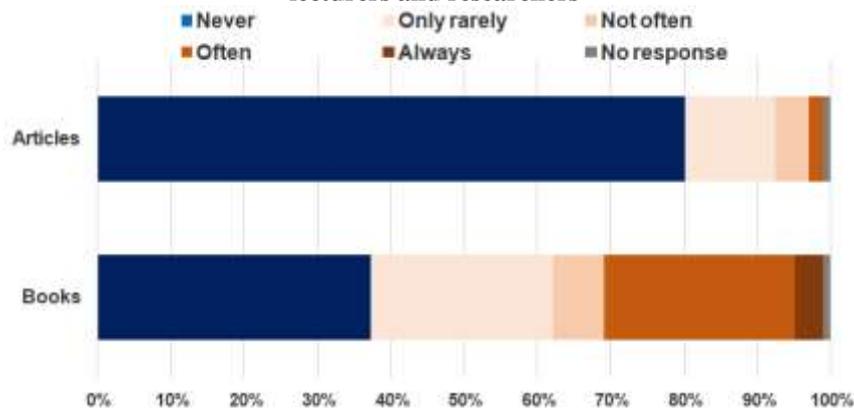

These results are firmly corroborated in the qualitative section of the study. All the interviewees stated that paying to publish a book was widespread and either they or their colleagues had personal experience of the practice. In addition, they explained that the cost of publishing books was usually covered by a publicly funded research project. They also acknowledged that this did not necessarily mean the books would be openly accessible, but the payment was made to cover some of the production costs in exchange for a specified number of copies. Some comments from the interviews are reproduced below:

> It's a fairly common practice. Funds from research projects are used to pay to publish books. And many publishers get practically all their business from this type of publication. P.Ph.1-i
>
> It's fairly common now among publishers… they'll only publish a collective volume with limited sales under these terms. Sl.Ph.2-i
>
> My experience is that with books associated with a research project, the books we write as a group, part of the publication costs were financed through the project. We always did that with departmental projects, at least in the field of political philosophy, and I think it's a fairly widespread practice. Sl.Ph,2-i
>
> My impression is that now nearly all SPI[3] publishers, the best ones, are all willing to listen to offers if a book is having difficulties getting sales. Sl.Ph.1-i

---

[3] SPI (Scholarly Publishers Indicators) is a system for evaluating and ranking scientific publishers in the field of humanities and social sciences in Spain. Available at http://ilia.cchs.csic.es/SPI/indexEn.html

> But I know the case, that is, in publishers where you have to pay to get published. Sl.E.2-i

On the question of the fees researchers paid, substantial differences emerged between article and book publications (Figure 3). Practically 80% of the articles published cost less than five hundred euros, whereas the fee was generally above a thousand euros (64%) for books. This difference seems logical, given the differences in the cost of publishing the two formats.

**Figure 3. Fees for publishing books and journal articles reported by university lecturers and researchers in philosophy and ethics in Spanish institutions.**

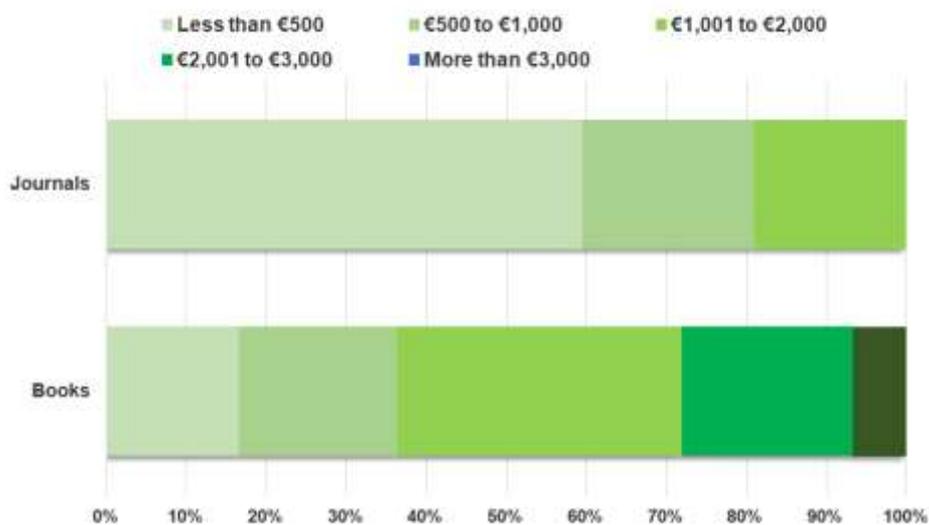

These data show a stark difference between publication practices, depending on whether the publication is an article or a book, reflecting two contrasting views of the publishing market for these two formats. In their responses, the researchers firmly rejected the pay-to-publish model for journal articles, which they considered a purely profit-driven activity. In contrast, there was clear support for paying to publish books, the hallmark of philosophical identity, due to the fragility (limited readership, low profitability) of this publishing market, which they considered must be preserved and supported.

*Rejection of the "commercialisation of publishing" in academic journals…*

Both the quantitative and qualitative results showed that paying to publish in journals is a relatively new phenomenon, and fairly unusual in the field of philosophy in Spain. However, despite its relatively recent appearance, the practice was heavily criticised in our study. The researchers regard it as an economic model driven by profit gained at the expense of others' efforts, but not really offering anything substantial in return. They do not even perceive the pay-to-publish fee as a way of subsidising OA to the article. As other studies have noted (Pinter, 2018, Dal-Re, 2019), the fees charged are generally considered exorbitant; furthermore, the researchers argued that it is inappropriate to allocate public money to private companies. Their comments are overwhelmingly condemnatory:

In the case of journals […], I think it's despicable to charge authors to maintain the system P.E.3-i

In the end what happens is that public money goes into private hands, that's what happens with the big publishers that control the most prestigious journals, the ones you have to pay to have your work published. P.E.1-d

What I think is bad, appalling, is the business that has developed around [the journals]. In other words, public money is financing an industry that isn't actually contributing anything. P.E.1-i

On the question of journals, the problem (which is also the subject of considerable international debate) is that a large proportion of the most prestigious journals are private. There's a whole publication industry there […] P.Ph.1-i

This means that they ask you for money to publish in a journal that, in addition, is disseminated through the internet (and so the production costs are very low), I think it's scandalous. […] You're paid to do research, you're paid so you can publish, and they make you pay if you want access to the actual research that you have done in your university. Sl.Ph.1-i

[…] I think the publishing conglomerates are abusing their power. In the end, what these huge companies are doing is exploiting your material (because you deliver them a document free of charge, the fruits of your labour, efforts and years of research) and on top of that you have to pay them. It's appalling. And recently, in academia, lecturers and researchers have to put up with everything. Sl.E.2-i

[…] journals that only publish on line, I understand there's business there. And the prices are very high. Sl.Ph.3-i

High impact journals, it's a market […]. And uncannily, almost all the top rated [journals] are English language. Philosophy produced in Spanish must be properly nurtured. Sl.Ph.11-s

*… But there is support for books (through payment) in a fragile philosophy publishing landscape in Spain*

The widespread rejection of paying to publish is reversed, however, where books are concerned. Various arguments are put forward in favour of payment, although the idiosyncratic nature of philosophy is always stressed. Financial support for publishers is defended and justified on the grounds that publishing academic philosophy is barely profitable. Some researchers also defend the cultural and social value of publishing books in the vernacular. They argue that this field should receive special protection and that it cannot be left to the dynamics of a market where academic publishing in the humanities has little chance of survival. These arguments are put forward by several researchers (Rf.E.1-i, Sl.Ph.1-i, Sl.Ph.3-i, Sl.Ph.2-i, P.E.2-i, L.Ph.1-i and L.Ph.2-I) as follows:

Some books hold their own in the market and others don't. I think that when a private publisher asks for help to publish a book, [the request] should be analysed, because in many cases these requests are made because the books are not viable on their own in the market. […] There are certain places that the market doesn't reach, and the State helps out because it considers this to be culturally relevant output. On the other hand, having to pay to publish a philosophy book is as old as philosophy itself. Nietzsche paid out of his own pocket to publish all his books, and in no way does this diminish the quality of a literary work. […] But even the best book, if Pierre Aubenque wanted to republish *Le problème de l'être chez Aristote* now, however good and serious a book it may be, no publisher would take it on because it's a limited market in business terms. I'm referring to monographs about authors, which do have a bigger readership. There's no telling what would happen if someone came along with a book like *The Order of Things*. There are books that, unless the author is so prestigious that their books sell regardless, it is very difficult to market them and make a profit. So we have two options. Either we ostracise all philosophical output that can't hold its own in the market, or, there are publishers that have found a niche in the market and publish work that falls between being able to hold its own in the market by selling 800 or 1000 copies, and work that needs support to be viable in this context. Over the years, I haven't seen this as just a predatory way for private publishers to get hold of public funds, but rather they are using public money to bring out books that wouldn't be published otherwise. In other words, they are collaborating (they aren't enemies) because it's difficult to keep a commercial publishing house afloat by publishing philosophy books with a certain level of exigency, academic quality or specialisation. Sl.Ph.1-i

[…] I don't consider it to be a bad thing at all. […] if publishers can't make a profit from publishing that book, because obviously it will be sold to a very limited public (libraries and specialists), I understand that we have to collaborate economically. Rf.E.2-i

[…] I think that if you have a project with funds and you can finance it, it's not a bad thing, so long as the amount they ask you for is reasonable. L.Ph1-i

I understand that in book publishing they sometimes ask you for money because dissemination is more difficult and has some costs. L.Ph2-i

Given that book sales are so low, I think it's a more or less acceptable kind of collaboration. Obviously, if the aim is to make a profit, the works we usually produce don't have very high commercial potential. So a prestigious publisher that produces a beautiful edition of your books, they ask you for money… but you also achieve one of the aims of the project. And moreover, the history of the group continues through its publications. I don't think that's a bad thing. Sl.Ph.2-i

I understand that marketing this type of product is complicated […]. I understand that for certain philosophical content there are publishers that receive funding. Sl.Ph3-i

Some respondents even argue that it is the responsibility of public administrations to safeguard book publication, and proposed a solution:

> What the Ministry for Education should have done […] is support university publishers, create an infrastructure for distribution, evaluation with editorial boards for the corresponding collections. […] The logical thing, when a young person does their thesis, would not be for the thesis to end up with a commercial publisher, they wouldn't be interested, but with a university publisher. That's what happens in the English speaking world. Where are many of the books on philosophy and other subjects published in the English speaking world? In university publishers. In Spain we have a deficit (the responsibility of public administrations) of support for university publishers as quality academic publishers. What would be logical would be for specialised academic work, which by definition will not have a large readership, to be published by university publishers. And commercial publishers would publish informative essays that have a wider audience, etc. PPh1-i

**Debates around pay to publish**

According to our informants, paying to publish raises doubts and suspicions about its potentially distorting effects on the integrity and quality of manuscript evaluation systems and equality of opportunity for authors faced with publication fees.

*Payment and questions of research quality*

For the philosophy and ethics researchers in our study, the effects on review processes of paying to publish are a key problem. Some argue that quality may be compromised because there is no guarantee of a rigorous, neutral review process with no conflicts of interest, as also noted in other studies (Gadagkar, 2016; Al-Khatib & da Silva, 2017). In other words, they believe that publication is conditioned exclusively by the payment, not by the virtue of the work itself, thereby corrupting the ethos of the peer review process. Some views on this subject were expressed as follows:

> The business of publishers charging… it's totally mercenary. Surely if without money there's no interest, why is there an interest when money is involved? That's the big question. The quality of what's published should be the priority, not money. The thing is, it's no less true that our output doesn't have a mass [readership] that will then defray the costs. P.E.3-i

> In effect, there is a paradox in the case of books: the publishers ask the researchers/authors for a fee to publish. This then corrupts the guarantee of the work's quality. If the publisher publishes the book simply because they have secured funding for printing, that's no guarantee of quality. We could say that it's the [evaluation] agencies themselves that in some way force authors to finance publication in commercial publishers, and what's more, with public funds, in many cases with money that comes from research funding. P.Ph1-i

> Paying corrupts the system. Now when you apply for research projects, a basic item of expenditure is payment to publish. Sl.E.4-d

> There are many mercenary publishers. Everyone knows that many of them have no future. And if you get 3000 or 5000 euros from a project, from wherever,

> they'll publish. I don't know where this is leading us, but not to a good place for sure. P.E.1-d

> Academic book and book chapter publication in Spain doesn't have a good reputation because it depends on payment, without any guarantee of open access to the publication. Sl.Ph.3-s

> This implies that there are no well-established quality criteria for what gets published, because it isn't the publisher's criterion that decides what is published [but the payment]. L.Ph2-i

In sum, paying to publish affects the other essential research criterion: the objective evaluation of publishing proposals. Paying to publish, the review process and quality are caught up in a complex relationship, especially in a fragile publishing system.

*Paying to publish and equal opportunities*

Another matter raised in our interviews is the concern about possible inequality in the resources available to researchers to meet publication fees, a subject previously raised in the literature (Mabe, 2004; Gadagkar, 2008; van Dalen, 2013; Leopold, 2014; Tzarnas & Tzarnas, 2015). Our respondents noted that some research lines may encounter greater difficulties in obtaining public funds, or that some sectors of the academic community, especially early career researchers, may find it more difficult to make such payments because they have not yet had the opportunity to access public or private funding. Indeed, other studies have reported costs as a major worry among young researchers struggling to pay APCs (Nicholas et al., 2019; Jamali et al., 2020; Rodríguez-Bravo & Nicholas, 2020; Nicholas et al. 2020).

Fewer informants in this study expressed their concerns in this respect but, significantly, they were two early career researchers, a group often affected by this form of inequality. They expressed their views as follows:

> But there are people who don't have [research] projects, or depending on other specific circumstances… I think it's more complicated, because does that imply that only people with projects or money can publish? This greatly limits who can actually publish. I think it is unfair to those who are left out of the academic system, or who are in precarious [conditions], and also those who are starting out. L.Ph1-i

> Now, what this [paying] means is that in the end certain people who have research projects and have extra funds are those who publish and it isn't accessible to everyone. L.Ph2-i

**On open access, public funding and research dissemination**

To find out about open access publication practices, we asked our researchers how many of their last ten publications were OA. The results showed a fairly high average of six publications. While 33.3% of the researchers reported that practically all their most recent papers were OA publications (between 8 and 10), only 15% had published two or less. What is the explanation for these figures? First, the Spanish central government committed to promote OA through Law 14/2011, which mandates OA publication of results from research mainly paid for with public funds (BOE, 2011). Second, there is a significant conviction among researchers of the importance of disseminating publicly funded work in OA publications without restrictions. Moreover, our results showed that the majority of Spanish philosophy and ethics academics (78%) held this view (Figure 4).

**Figure 4. Opinion of university lecturers and researchers in philosophy and ethics on the obligation to publish in open access all research financed with public funds**

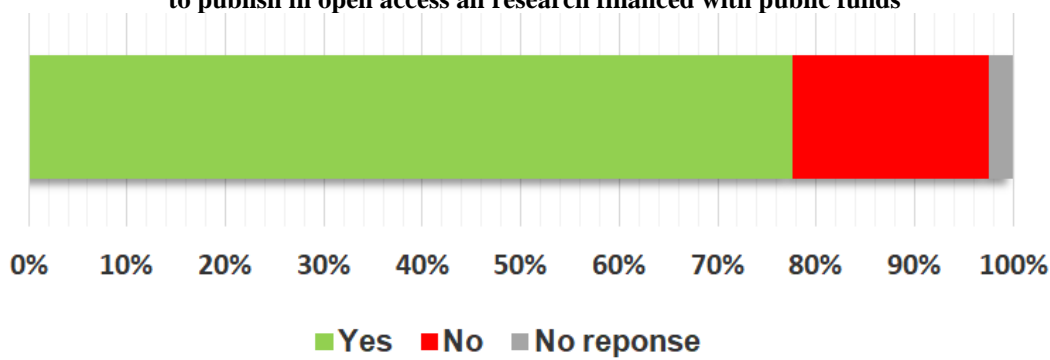

In the qualitative section of the study, the researchers were also in favour of OA dissemination of all research financed with public funds. Indeed, they considered it as a logical obligation. Some of their views are reflected in the following comments:

> All research carried out with financing from the Ministry or the European Research Council should be published in open access publications because it's funded with public money. Rf.E.1-i

> Public institutions must meet the costs of maintaining and updating open access dissemination platforms, without depending on groups where there may be conflicts of interest or where quality criteria may be conditioned by the goal of obtaining profits. L.Ph.5-s

> As for journal articles, I think it's absolutely right to demand that work financed with public funds be available in open access. Sl.Ph.1-I

> I always submit to open access journals. When I'm deciding where to publish, I always check whether it's open access. L.Ph.1-i

Finally, a third possible explanation for the high percentage of OA publications lies in the publishing landscape of Spanish journals. As mentioned earlier, they are supported by public funds, and no payment is required either to publish in them or to read their contents. The qualitative results of our study in particular reveal that researchers tend to associate the concept of OA as being free of costs. For many respondents, the two concepts were synonymous, as reflected in the following statements:

> The results of any research, especially if carried out in public institutions and/or financed with public funds, must have free and open access… P.Ph.3-s

> Especially in research, ideally it should be as accessible as possible to everyone and should exclude any payment for the objective of the research to succeed. The more accessible and simple the better, because it will advance research. L.Ph.2-i

> The space for publicly funded publications (especially university journals) must be restored, and no one should have to pay for access. P.Ph.1-i

In sum, support for OA is conditional on a specific model, based on public subsidies for journals and free access for both authors and readers.

**Discussion/conclusions**

The pay-to-publish model is met with rejection (at least *a priori*) in the Spanish philosophy research community. Indeed, not having to pay to publish is the third most important criterion for researchers when they are deciding where to publish their work. The pay-to-publish model is not well established in philosophy journals, although this is not the case in book publication. This is unsurprising if we take into account that in the Spanish publishing context, practically 90% of humanities journals are freely accessible (Melero, 2017). Previous studies have shown that APCs are uncommon in philosophy, and account for just 3% of the publications listed in the Directory of Open Access Journals (Kozak & Hartley, 2013). Other studies have found that 87.3% of researchers paid no fee to publish their latest OA article in 2010, and 83.3% of researchers in 2016 (Ruiz-Pérez, 2017). Our data coincide in that a large majority of the respondents (80%) never had to pay a fee, and another significant percentage (12%) paid only infrequently. In contrast, paying to publish books is a widespread practice in the areas covered in our study.

    A central argument noted in the researchers' perceptions is that paying to publish derives from a commercial model of scientific dissemination. In general, the researchers associate paying to publish articles with an industry that aims to make disproportionate profits, as noted by other authors (Buranyi, 2017; Dal-Re, 2019). This negative view is tempered, however, in the case of the book publishing industry, where the researchers put forward arguments in favour of publication fees. It is worth noting that in the areas of philosophy and ethics in Spain, the pay-to-publish debate is grounded in justifications that diverge from the usual arguments. Paying to publish is often justified as the way to ensure open access to published work. Those funding the research cover the author's publication costs so their work appears in OA publications (Truth, 2012; Björk & Solomon, 2015, Björk, 2017; Severin et al., 2018; Edelmann & Schoßböck, 2020). This is not the case here, however, where the option is presented as a fee to support and guarantee the 'survival' of the book, but not to provide a free, OA product. Although this phenomenon has been noted previously (Giménez-Toledo, Tejada-Artigas & Borges-De-Oliveira, 2019), it has not been studied in depth. One reason that might explain this position is that books, as studied here in the case of the humanities in Spain (Giménez-Toledo, 2016), are seen as a hallmark of identity by researchers in the areas of philosophy and ethics (Delgado-López-Cózar, Feenstra & Pallarés, 2020). A second reason for their willingness to pay is that they consider the precarious publishing industry deserves support through publicly funded research projects, in the same way as other cultural sectors are supported in Spain.

    However, the researchers in our study were not unanimous in their views on paying to publish, as reflected in the debates on the negative impact the practice can have on the review process, academic quality, and equality of opportunities, as also noted in previous studies and forums (Mabe, 2004; van Dalen, 2013; Leopold, 2014; Gadagkar, 2008; Tzarnas & Tzarnas, 2015; Gadagkar, 2016; Al-Khatib & da Silva, 2017). Some of the respondents expressed concern that the quality of published work may be under threat if publishers face a conflict of interest between publishing for mere economic profit and offering real quality, or that the process of review and critical, rigorous and neutral evaluation of their work may be compromised. The question some of them raise, therefore, is whether a review process really takes place when payment is

involved. They also note inequality of opportunities as a potential perverse effect, since not everyone is in the same position to obtain financial resources.

In turn, the researchers were in favour of OA dissemination for publicly funded research, in line with findings from other studies in Spain focusing on early career researchers (Rodríguez-Bravo & Nicholas, 2019; Rodríguez-Bravo & Nicholas, 2020). This attitude is confirmed by the researchers' publication practices, as reflected in the significant percentage of their recent publications that came out in OA formats. It should be noted, however, that when we asked the researchers about the criteria they followed when selecting a journal or publisher for their work, OA was ranked the second least relevant. It is important to stress, therefore, that so much of the ethics and philosopher researchers' work appears in OA publications because of the current journal publishing landscape. There also appears to be some confusion among these researchers over the meaning of OA, in that they identify and understand it as a publication model that is free for both authors and readers and that is financed with the support of public funds. Future research could usefully extend these data and examine, for example, the extent of Spanish philosophy researchers' knowledge about OA options, and their opinions about the various financing models. For the moment, we have uncovered some interesting aspects regarding pay to publish and OA in philosophy and ethics in Spain, how widespread the practices are, and the debates they have raised.

**Ethical Statement**

**Acknowledgements**

The authors would like to thank Daniel Pallarés-Domínguez for his help in the data collection process. We would also like to thank the three main Spanish philosophy and ethics associations, AEEFP, SAF and REF, for their collaboration during the research process and their endorsement of the data collection report.